\documentclass{article}
\pdfoutput=1
\usepackage{graphicx,amsmath,amssymb,epsf} \usepackage{latexsym,bm,
  slashed} \usepackage{xcolor} \definecolor{dark}{rgb}{0.10,0.2,0.3}
\definecolor{magenta}{rgb}{0.7,0.1,0.3}
\definecolor{purpure}{rgb}{0.5,0.15,0.3}
\usepackage[font=small,format=plain,labelfont=bf,up,textfont=it,up]{caption}
\usepackage{hyperref, cite} \hypersetup{colorlinks, citecolor=blue,
  filecolor=blue, linkcolor=magenta,
  urlcolor=purpure,hyperfootnotes=true,pdftex} 
  
\newcommand{\tr}{{\rm tr}}

\title{An effective field theory approach for electroweak interactions\\ in the high energy limit } 
\author{Melina G{\' o}mez Bock$^1$, Martin Hentschinski$^1$, Agust{\' \i}n Sabio Vera$^2$\\ 
\\
\small
$^1$ Departamento de Actuaria, F{\' \i}sica y Matem{\' a}ticas,\\ \small Universidad de las Am{\' e}ricas Puebla, \\\small Santa Catarina M{\'a}rtir, 72820 Puebla, M{\'e}xico.\\
\small $^2$ Instituto de F{\' \i}sica Te{\' o}rica UAM/CSIC, Nicol{\'a}s Cabrera 15\\ \small
\& Universidad Aut{\' o}noma de Madrid, E-28049 Madrid, Spain.} 

\begin{document} 

\maketitle 

\begin{abstract}
We present an effective action for the electroweak sector of the Standard Model valid for  the calculation of scattering amplitudes in the high energy (Regge) limit. Gauge invariant Wilson lines are introduced to describe reggeized degrees of freedom whose interactions are generated by effective emission vertices. From this approach previous results at leading logarithmic accuracy for electroweak boson Regge trajectories are reproduced together with the corresponding interaction kernels. The proposed  framework lays the path for calculations at higher orders in perturbation theory. 
\end{abstract}

\section{Introduction}

The high energy limit of vector boson scattering in the electroweak sector of the Standard Model has been investigated in detail in~\cite{Bartels:2006kr}. The results are obtained in the multi-Regge limit of the theory and parallel those in QCD within the BFKL approach~\cite{Lipatov:1976zz,Fadin:1975cb,Kuraev:1976ge,Kuraev:1977fs,Balitsky:1978ic,Balitsky:1979ns}.  In~\cite{Bartels:2006kr} the main result is the derivation of an integral equation suited for the description of vector particle scattering in the vacuum exchange channel.  The emerging picture is that of interacting reggeized charged particles and non-reggeizing neutral vector bosons. The consistency of these effective high energy degrees of freedom is sustained by the isospin-1 exchange bootstrap equations derived in~\cite{Bartels:2006kr}. 

Within this framework, our contribution in the present work is to offer an effective action useful to handle these high energy fields and evaluate efficiently scattering amplitudes for the interaction among electroweak vector bosons. We show how to reproduce the reggeization scheme above mentioned together with the calculation of effective production vertices needed to construct the integral evolution equations presented in~\cite{Bartels:2006kr}. Our calculational framework is based on that proposed by Lipatov~\cite{Lipatov:1995pn,Lipatov:1996ts,Antonov:2004hh} and our own work in the QCD context~\cite{Hentschinski:2008rw,Hentschinski:2011tz,Hentschinski:2011xg,Chachamis:2012gh,Chachamis:2012mw,Chachamis:2012cc,Chachamis:2013hma,Hentschinski:2014lma,Hentschinski:2014bra,Hentschinski:2014esa}. 

One of the virtues of this effective action approach is to set up a powerful tool to evaluate higher order corrections in the electroweak high energy effective theory. Special attention is put to the role of the scalar Higgs boson in the unitarity properties of these amplitudes and its contribution to the Pomeranchuk singularity governing the growth with energy of the total cross section.  Our results should have an interesting interplay with the electroweak parton distributions discussed 
in~\cite{Ciafaloni:2008cr}. They can be of importance also in the evolution of parton distribution and fragmentation functions up to scales beyond the electroweak scale (where the $SU(2) \times U(1)$ symmetry is unbroken)~\cite{Bauer:2017isx,Bauer:2017bnh,Bauer:2018xag,Manohar:2018kfx,Fornal:2018znf}. Lipatov's high energy effective action has had an abundant number of applications in QCD, see, {\it e.g.}~\cite{Nefedov:2013ywa,Bondarenko:2018kqs,Hentschinski:2018rrf,Karpishkov:2017kph,Braun:2017qij,Braun:2016sij,Nefedov:2014qea,Nefedov:2020ecb,Bondarenko:2019llt,Braun:2020thc}, in supersymmetric theories~\cite{Bartels:2008ce,Bartels:2008sc} and 
gravity~\cite{Lipatov:1982vv,Lipatov:1991nf,Lipatov:2011ab,Bartels:2012ra,SabioVera:2019edr,SabioVera:2019jqe}. 

In the next section we introduce the effective action for a general non-abelian $SU(N)$ theory where the associated gauge bosons are known to reggeize in the leading logarithmic approximation (LLA)~\cite{Grisaru:1973vw,Lipatov:1976zz,Fadin:1975cb,Kuraev:1976ge,Kuraev:1977fs,Balitsky:1978ic,Balitsky:1979ns}. In Section 3 we extend the formalism to the unbroken $SU(2) \times U(1)$ case taking into account that the abelian gauge boson of QED does not reggeize in the LLA. We then introduce a vacuum expectation value for the Higgs field to break the symmetry and develop the high energy effective action for the different kinematical components of the gauge fields, treating the charged and neutral sectors on the same footing. In Section 4 we extract the associated Feynman rules which will allow us to evaluate the electroweak gauge bosons Regge trajectories in 
Section 5 and the effective production vertices, needed to construct the integral equations proposed in~\cite{Bartels:2006kr}, in Section 6. We draw some Conclusions after this. 

\section{Effective action for the $SU(N)$ sector}

The analysis of the $SU(N)$ sector is equivalent to that of QCD without fermions, which can be neglected in the LLA. This is also relevant in the electroweak case since in the Higgs sector of the Standard Model, with a spontaneously broken  $SU(2)$  symmetry, the non-abelian gauge bosons also reggeize. The corresponding Lagrangian reads
\begin{eqnarray}
  {\cal L}_{\rm SU(N)}[v_\mu]
  &=&   \frac{1}{2} \tr \left[G_{\mu\nu} G^{\mu\nu} \right] \, , 
\end{eqnarray}
where  $D_\mu = \partial_\mu + g v_\mu$, $G_{\mu\nu} =  [D_\mu, D_\nu]/g$, and the vector fields $v_\mu = -i t^a v_\mu (x)$ are considered as anti-hermitian matrices acting in the $SU(N)$ fundamental representation with generators $t^a$ fulfilling $\tr (t^a t^b) = \delta^{ab} / 2$. 

\begin{figure}[h]
\vspace{-5cm}
  \centering
  \includegraphics[width=10cm]{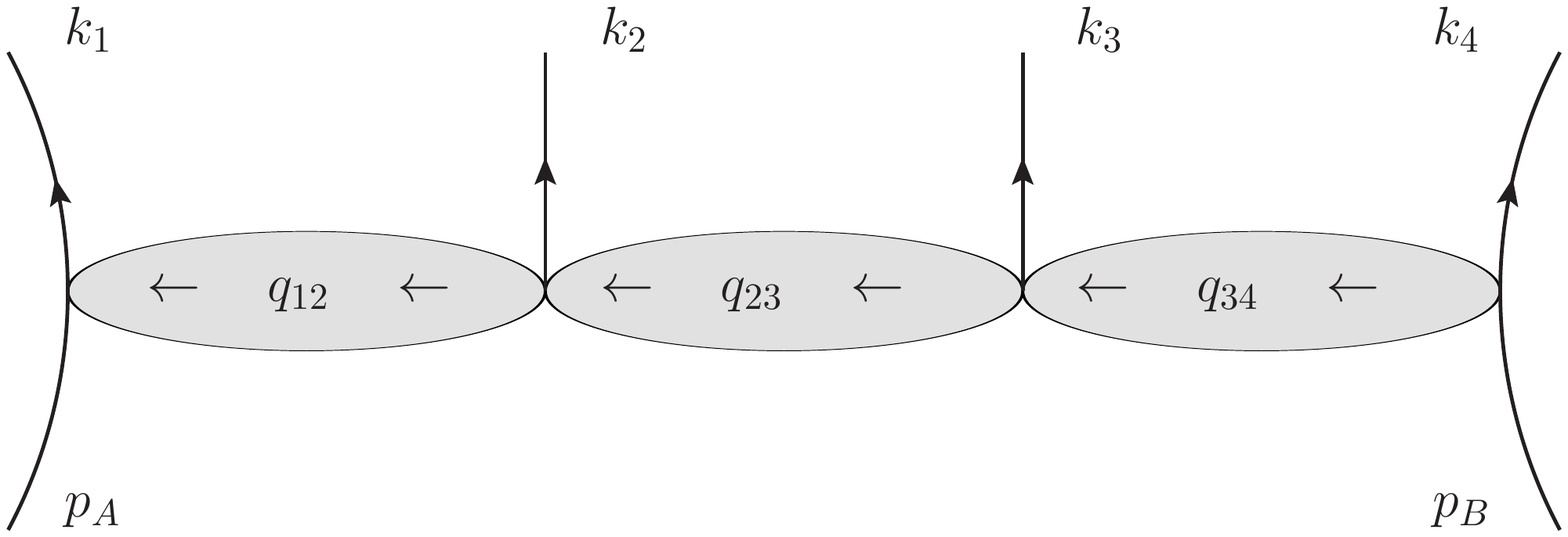}
  \vspace{-5cm}
  \caption{Kinematics of the $2 \to 4$ scattering process.}
  \label{fig:mrk}
\end{figure}

To implement the effective description of multi-Regge kinematics (MRK), which is the dominant region governing high energy scattering in the leading logarithmic approximation (LLA), we first present Fig.~\ref{fig:mrk} as a reference for the momenta conventions (in this case for a $2 \to 4$ process). 
Clusters of particles are represented by emissions with momenta $k_i$ in the graph. In the Regge limit we have $s=(p_A+p_B)^2 = (2E)^2 \gg -t$. In the LLA only one particle is allowed per cluster, with two in the next-to-leading order approximation (NLLA) and so on. Each cluster is emitted at rapidity 
\begin{align}
  \eta_i &= \frac{1}{2} \ln \frac{k_i^+}{k_i^-}
\end{align}
where $ p_{A,B}^\pm n_\pm/2 = p_{A,B}$ and $k_\pm = k \cdot n_\pm = k^\mp = k \cdot n^\mp$; furthermore $(n^\pm)^2 = 0$ and $n^\pm \cdot n^\mp = 2$.  In the LLA we resum 
$g^2 \ln{(s/s_0)}$ terms (with $s_0$ a suitable, process dependent reference scale) and the exchange in the $t$-channel of vector bosons is the dominant contribution with fermions and scalars being relevant only in the NLLA and beyond. 

The target is to describe in a gauge invariant way the interaction of particles locally in rapidity within each cluster and non-locally with reggeized vector bosons exchanged 
among different clusters. The separation into different modes is realised using the following decomposition into local and non-local components with each rapidity gap sector (with no emissions) denoted by $s$, {\it i.e.}
\begin{eqnarray}
  v_\mu(x) &=& \sum_{s} v^{(s)}_\mu(x) ~=~ \sum_{s} \left [ V_\mu^{{\rm local} (s)}(x) +  \frac{n^+}{2} V_+^{\eta > \eta(s)} +   \frac{n^-}{2} V_-^{\eta < \eta(s)} \right] \, .
\end{eqnarray}
This separation translates into $S^{(s)} = \int d^4 x {\cal L}_{SU(N)} [v^{(s)}_\mu (x)]$, the corresponding effective action for each local sector, where bilinear terms generating the transition from a single local into a single non-local field do not contribute. This cancelation can be made explicit in the effective action introducing the Sudakov decompositions 
\begin{eqnarray}
v^{(s)}_\mu (x) &=& v^{(s)}_{-} (x) \frac{n^-_\mu}{2}
+  v^{(s)}_{+} (x) \frac{n^+_\mu}{2}  +  v^{(s)}_{\perp} (x) \, ,\\
V_\mu^{{\rm local} (s)}(x) &=& V_-^{{\rm local} (s)}(x) \frac{n^-_\mu}{2}
+  V_+^{{\rm local} (s)}(x)  \frac{n^+_\mu}{2}  +  V_\perp^{{\rm local} (s)}(x) \, ,
\end{eqnarray}
and directly subtracting the bilinear terms using
\begin{align}
  S^{(s)} = \int d^4 x {\cal L}_{SU(N)} [v^{(s)}_\mu (x)]
  & + \int d^4 x \, \tr \left[\left(v_-  - V_-^{\eta <  \eta(s)} \right) \partial^2
 V_+^{\eta >  \eta(s)}
 \right] \notag \\
& + \int d^4 x \, \tr \left[\left(v_+  - V_+^{\eta >  \eta(s)} \right) \partial^2
 V_-^{\eta <  \eta(s)} \right] \, ,
\end{align}
where
\begin{eqnarray}
\label{VsLocal}
v^{(s)}_{-}  - V_-^{\eta <  \eta(s)} ~=~ V_-^{{\rm local} (s)} \, \, \, , \, \, \, 
v^{(s)}_{+}  - V_+^{\eta >  \eta(s)} ~=~ V_+^{{\rm local} (s)}  \, .
\end{eqnarray}

The next step is to replace the fields $ V_+^{\eta > \eta(s)} , V_-^{\eta < \eta(s)} $ by two new fields $A_\pm$ which are invariant under local gauge transformations and subject to the constraints $\partial_+ A_- = 0 = \partial_- A_+$ (for convenience we remove the $s$ index from now on) to impose the $\eta < \eta(s)$ $(\eta > \eta(s))$ rapidity modes separation.  We can therefore write
\begin{align}
   S = \int d^4 x {\cal L}_{SU(N)} [v_\mu (x)]& + \int d^4 x \, \tr \left[\left(v_-  - A_- \right) \partial^2
 A_+
 \right]  + \int d^4 x \, \tr \left[\left(v_+  - A_+ \right) \partial^2
 A_-\right] \, .
 \label{SeffSUN}
\end{align}
The fields $A_\pm$ correspond to the reggeized vector bosons in the theory. 
It is finally needed to introduce the replacement 
\begin{align}
  v_\pm(x) \to v_\pm(x) U[v_\pm(x)] = -\frac{1}{g} \partial_\pm U[v_\pm(x)]
\end{align}
where the path-ordered Wilson line 
\begin{align}
  U[v(x)] & = \mathcal{P} \exp \left(-\frac{g}{2} \int_{-\infty}^{x^\pm} d {z}^\pm v_\pm(z) \right)
\end{align}
ensures the full gauge invariance in the effective action of Eq.~(\ref{SeffSUN}). 
 
\section{Effective action for the electroweak interaction}

When constructing the high energy effective action for the electroweak sector of the Standard Model it is important to take into account that not all vector fields do 
actually reggeize. Besides this point, the procedure will be similar to the previous section and the vector fields $V_\mu = W_\mu, B_\mu$ of the $SU(2) \times U(1)$ theory will be 
expanded into local and non-local rapidity modes. 

We will subtract those contributions  coupling local and non-local fields by means of an induced term in the effective action. The  
non-local fields will be promoted to gauge invariant high energy fields where not all of them will correspond to reggeized fields, this is distinctly different to the QCD case where both quarks and gluons do reggeize in the high energy limit. Finally, suitable path-ordered Wilson lines will be introduced in order to complete the construction of a gauge invariant 
high energy effective action for the electroweak sector in the Standard Model. 

To implement this program, let us first introduce the $SU(2) \times U(1)$ gauge fields as
\begin{align}
w_\mu(x) &= -i t^a W_\mu^a(x) \, , & b_\mu(x) & = -i B_\mu(x)
\frac{1}{2}\begin{pmatrix}
  1 & 0 \\ 0 & 1
\end{pmatrix} \, ,
\end{align}
where $t^a$ are the generators of the fundamental representation of SU$(2)$
with $\tr(t^at^b)=1/2$. The covariant derivatives read
\begin{align}
 D[w]_\mu & = \partial_\mu + g w_\mu   \, ,  &     D[b]_\mu & = \partial_\mu + g' b_\mu\, ,
\end{align}
and the field strength tensors are
\begin{align}
  F_{\mu\nu}^a[W] & = \partial_\mu W_\nu^a - \partial_\nu W_\mu^a + g
                    f^{abc} W^b_\mu W^c_\nu \, ,
&
F_{\mu\nu}[B] & = \partial_\mu B_\nu - \partial_\nu B_\mu \, , \notag \\
  F_{\mu\nu}^a[W]t^a & = \frac{1}{-ig}[D[w]_\mu, D[w]_\nu]  \, ,
   & 
  F_{\mu\nu}[B] \frac{1}{2}\begin{pmatrix}
  1 & 0 \\ 0 & 1 \end{pmatrix}& = \frac{1}{-ig'}[D[b]_\mu, D[b]_\nu] \, ,
\end{align}
where
\begin{align}
 F_{\mu\nu}[w] & \equiv \frac{1}{g}[D[w]_\mu, D[w]_\nu] \, , &  F_{\mu\nu}[b] & \equiv \frac{1}{g'}[D[b]_\mu, D[b]_\nu]  \, ,
\end{align}
are understood as $2\times 2$ matrices. The kinetic terms for the gauge fields are
\begin{align}
    -\frac{1}{4} F[W]^a_{\mu\nu}F[W]^{a,{\mu\nu}} 
 & -\frac{1}{4} F[B]_{\mu\nu}F[B]^{{\mu\nu}}  = \frac{1}{2}\tr\left( \frac{1}{2} F[w]_{\mu\nu}F[w]^{{\mu\nu}} +  F[b]_{\mu\nu} F[b]^{\mu\nu} \right) \, .
\end{align}
The covariant derivative associated to the Higgs field considered as a  $SU(2)$ scalar complex doublet, $\phi$, reads
\begin{align}
  D_\mu \phi(x) & = \left(\partial_\mu + g w_\mu + g' b_\mu \right)\phi(x) \, .
\end{align}
Together with the Higgs potential $V(\phi) = \mu^2 \phi^\dagger \phi - \lambda (\phi^\dagger \phi)^2$ this yields the well-known electroweak Lagrangian (neglecting the coupling to fermions):
\begin{align}
  \mathcal{L}_{\rm EW}[w_\mu, b_\mu, \phi] & = 
                \frac{1}{2}\tr\left( F[b]_{\mu\nu} F[b]^{\mu\nu} 
                +
                \frac{1}{2} F[w]_{\mu\nu}F[w]^{{\mu\nu}} \right)
                +
                \left|  D_\mu \phi(x)\right|^2 + V(\phi) \, .
\end{align}

The separation into different rapidity modes can be achieved using 
\begin{align}
  w_\mu(x) & = \sum_{s} \left [ W_\mu^{local, s}(x) +  \frac{n^+}{2} W_+^{\eta > \eta(s)} +   \frac{n^-}{2} W_-^{\eta < \eta(s)} \right]  \notag \, , \\
  b_\mu(x) & = \sum_{s} \left [ B_\mu^{local, s}(x) +  \frac{n^+}{2} B_+^{\eta > \eta(s)} +   \frac{n^-}{2} B_-^{\eta < \eta(s)} \right] \, , 
\end{align}
where
\begin{align}
  \partial_- W^{\eta > \eta (s)}_+(x) & = 0 =  \partial_+ W^{\eta < \eta (s)}_-(x) \, ,
&
\partial_- B^{\eta > \eta (s)}_+(x) & = 0 =  \partial_+ B^{\eta < \eta (s)}_-(x) \, .
\end{align}

Taking into account that the Higgs field is always local in the rapidity mode expansion, before spontaneous symmetry breaking the coupling between a local and a non-local field lies within the kinetic terms of the gauge fields. The situation is more complicated when the Higgs field acquires a vacuum expectation value of the form
\begin{align}
  \phi_0 & = \langle \phi(x)\rangle = \frac{1}{\sqrt{2}}
           \begin{pmatrix}
             0 \\ v
           \end{pmatrix}
\end{align}
since in the kinetic term of the Higgs field we find 
\begin{align}
 \tr \left[ w_\mu  g^2M^2  w_\mu   +b_\mu {g'}^2 M^2  b_\mu + gg' \left(w_\mu M^2 b_\mu + b_\mu M^2 w_\mu \right)\right]
\end{align}
where
\begin{align}
  \phi_0 \otimes \phi_0^\dagger & =  \frac{v^2}{2}
                                \begin{pmatrix}
                                  0 & 0 \\ 0 & 1
                                \end{pmatrix} = M^2 \, .
\end{align}
This generates local to non-local field transitions after kinematical separation in rapidity of the electroweak boson fields of the form 
\begin{align}
\label{HiggsSubtraction}
                                 &  \frac{1}{2}\tr \bigg[
                                  W^{local}_+  g^2M^2 W_-^{\eta < \eta(s)} 
                                  + 
                                  W_-^{\eta < \eta(s)}  g^2M^2  W^{local}_+ 
                                   +
                                  B^{local}_+  {g'}^2M^2 B_-^{\eta < \eta(s)} 
                                  + 
                                  B_-^{\eta < \eta(s)}   {g'}^2 M^2  B^{local}_+
                                  \notag \\ & +
                                  gg'\left(
                                  W^{local}_+ M^2 B_-^{\eta < \eta(s)} 
                                  + 
                                  B_-^{\eta < \eta(s)}  g^2M^2  W^{local}_+
                                  +
                                  B^{local}_+ M^2 W_-^{\eta < \eta(s)} 
                                  + 
                                  W_-^{\eta < \eta(s)}  g^2M^2  B^{local}_+
                                  \right) \notag \\
&
                                   + \{ ``+" \leftrightarrow ``-" , ``>" \rightarrow ``<"\} \bigg] \, .
\end{align}

As explained in the previous section in the pure $SU(N)$ theory, these terms need to be 
subtracted from the original electroweak effective action in order to correctly capture the multi-Regge limit of scattering amplitudes. Naming Eq.~(\ref{HiggsSubtraction}) as ``Higgs subtraction" term, for the effective action valid in the multi-Regge limit $ {S}^{\rm EW}_{\rm eff}$ we would have
\begin{eqnarray}
  {S}^{\rm EW}_{\rm eff}   &=& \int d^4 x \,  \mathcal{L}_{\text{EW}}[w_\mu, b_\mu, \phi ]
  ~+~ \int d^4 x \, \tr \left[\left(w_-  - W_-^{\eta <  \eta(s)} \right) \partial^2 W_+^{\eta >  \eta(s)} \right] \notag \\
 &+& \int d^4 x \, \tr \left[\left(b_-  - B_-^{\eta <  \eta(s)} \right) \partial^2
 B_+^{\eta >  \eta(s)} \right]   - \text{Higgs subtraction}  \notag\\
 &+& \{ ``+" \leftrightarrow ``-" , ``>" \rightarrow ``<"\} \, .
\end{eqnarray}
As in Eq.~(\ref{VsLocal}) we now write $W^{local}_- = w_-  - W_-^{\eta <  \eta(s)} $ and $W^{local}_+ = w_+  - W_+^{\eta >  \eta(s)}$ with equivalent expressions for the $B$ field. The next step is to promote the non-local fields to gauge invariant high energy fields $A_\pm^{(w)}$ and $A_\pm^{(b)}$ subject to the kinematical constraints $\partial_\mp  A_\pm^{(w)} = 0 = \partial_\mp  A_\pm^{(b)}$. We therefore have an effective action written as the sum of the usual Standard Model electroweak action plus additional ``induced" terms (using Lipatov's notation) where the Lagrangian for  high energy fields reads
\begin{eqnarray}
 \mathcal{L}_{\rm induced}  &=&  \frac{1}{2} \tr \bigg[ (w_+ - A_+^{(w)})  \left(\partial^2  + {g^2}M^2\right) A_-^{(w)}  +  A_-^{(w)} \left(\partial^2  + {g^2}M^2\right) (w_+ - A_+^{(w)})  
\notag \\
&+& (b_+ - A_+^{(b)}) \left(\partial^2  +{g'}^2  M^2\right) A_-^{(b)} 
+  A_-^{(b)} \left(\partial^2  +{{g'}^2} M^2\right) (b_+ - A_+^{(b)})
\notag \\
&+& {gg'} \bigg\{(w_+ - A_+^{(w)}) M^2 A_-^{(b)}  +  A_-^{(b)}  M^2 (w_+ - A_+^{(w)}) 
\notag \\
&+& (b_+- A_+^{(b)}) M^2 A_-^{(w)} + A_-^{(w)}  M^2  (b_+- A_+^{(b)})
\bigg\}
\bigg] + \{``+ " \leftrightarrow ``-" \} \, .
\end{eqnarray}

This action is not yet fully gauge invariant since, even though the abelian $b$ field is invariant under local gauge transformations, the non-abelian $w$ field is not. Similarly to the $SU(N)$ case in the previous section it is needed to replace $w_\pm$ by a path-ordered Wilson line
\begin{align}
\label{TWilsonLine}
  w_\pm \to -\frac{1}{g}\partial_\pm U[w_\pm(x)] \equiv T[w_\pm] \, .
\end{align}
One should therefore write
\begin{eqnarray}
 \mathcal{L}_{\rm induced}  &=&   \frac{1}{2}\tr \bigg[ (T[w_+] - A_+^{(w)})  \left(\partial^2  + {g^2}M^2\right) A_-^{(w)}  +  A_-^{(w)} \left(\partial^2  + {g^2}M^2\right) (T[w_+] - A_+^{(w)})  
\notag \\
&+& (b_+ - A_+^{(b)}) \left(\partial^2  +{{g'}^2} M^2\right) A_-^{(b)} 
+  A_-^{(b)} \left(\partial^2  +{{g'}^2} M^2\right) (b_+ - A_+^{(b)})
\notag \\
&+& {gg'} \bigg\{(T[w_+] - A_+^{(w)}) M^2 A_-^{(b)}  +  A_-^{(b)}  M^2 (T[w_+] - A_+^{(w)}) 
\notag \\
&+& (b_+- A_+^{(b)}) M^2 A_-^{(w)} + A_-^{(w)}  M^2  (b_+- A_+^{(b)})
\bigg\} \bigg]  + \{``+" \leftrightarrow ``-" \} \, .
\end{eqnarray}

\section{Effective Feynman rules}

To study the high energy behaviour of electroweak scattering amplitudes in the 
multi-Regge limit two ingredients are needed in the LLA: effective emission vertices and electroweak boson Regge trajectories. We will now focus on identifying the induced effective vertices generated by the high energy action which will actually contribute in the calculation of those two fundamental components of the effective field theory. 

In order to calculate the electroweak gauge bosons Regge trajectories
it is needed to first find the propagator of the high energy fields,
{\it i.e.} to evaluate the two-point vertex of a $A_+$ and a $A_-$
field, $V_{A_+,A_-}(q^2)$. To this end we express the quadratic term
of the induced Lagrangian in the basis of mass eigenstates (with
$M_Z^2 = \frac{M_W^2}{\cos^2{(\theta_W)}} = \frac{ v^2}{4} (g^2+{g'}^2
)$ where $\theta_W$ is Weinberg's angle), {\it i.e.}
\begin{eqnarray}
 \mathcal{L}_{\rm induced}^{\rm quadratic}  &=&  - \frac{1}{2} \bigg\{
\left(W_+^+ - A_+^{(W^+)} \right)\left(\partial^2 + M_W^2 \right)A_-^{(W^-)}
+
\left(W_+^- - A_+^{(W^-)} \right)\left(\partial^2 + M_W^2 \right)A_-^{(W^+)}
\notag \\
&+&  \left(Z_+ - A_+^{(Z)} \right)\left(\partial^2 + M_Z^2 \right)A_-^{(Z)}
+
\left(C_+ - A_+^{(C)} \right) \partial^2 A_-^{(C)}
\bigg\} + \{``+" \leftrightarrow ``-" \} \, ,
\end{eqnarray}
where $C_\mu$ is the photon field  and $R_\pm$ denotes $n_\pm^\mu R_\mu$ (with $R=W^\pm,Z,C$).  The propagator in momentum space can then
be obtained making use of
\begin{align}
  G_{A_+A_-}(q^2) & = \frac{-1}{V_{A_+A_-}(q^2)}\bigg|_{q^2 \to q^2 +
                    i \epsilon}
\end{align}
where the vertices receive both contributions from the conventional electroweak Lagrangian and the induced Lagrangian.  The full set of relevant vertices and propagators for the high energy fields (not reggeized degrees of freedom yet) reads 
\begin{align}
  \label{eq:propR0}
  V_{A_+^{(W^+)}A_-^{(W^+)}}(q^2) & = \frac{i}{2} (-q^2 + M_W^2) \, , & 
 G_{A_+^{(W^+)}A_-^{(W^-)}} & = \frac{-1}{V_{A_+^{(W^+)}A_-^{(W^-)}} (q^2 + i\epsilon)} = \frac{-2i}{q^2 - M_W^2 + i \epsilon} \, , \notag \\
 V_{A_+^{(W^-)}A_-^{(W^+)}}(q^2) & = \frac{i}{2} (-q^2 + M_W^2) \, , &
  G_{A_+^{(W^-)}A_-^{(W^+)}} & = \frac{-1}{V_{A_+^{(W^-)}A_-^{(W^+)}} (q^2 + i\epsilon)} = \frac{-2i}{q^2 - M_W^2 + i \epsilon} \, ,\notag \\
 V_{A_+^{(Z)}A_-^{(Z)}}(q^2) & = \frac{i}{2} (-q^2 + M_Z^2) \, , & 
 G_{A_+^{(Z)}A_-^{(Z)}} & = \frac{-1}{V_{A_+^{(Z)}A_-^{(Z)}} (q^2 + i\epsilon)} = \frac{-2i}{q^2 - M_Z^2 + i \epsilon} \, ,\notag \\
 V_{A_+^{(\gamma)}A_-^{(\gamma)}}(q^2) & = \frac{i}{2} (-q^2) \, , &  G_{A_+^{(\gamma)}A_-^{(\gamma)}} & = \frac{-1}{V_{A_+^{(\gamma)}A_-^{(\gamma)}} (q^2 + i\epsilon)} = \frac{-2i}{q^2 + i \epsilon} \, .
\end{align}
As the next step we need to determine the first induced vertex, which corresponds to the interaction of three fields in the induced Lagrangian. With this target in mind, it is convenient to switch to a different $SU(2)$ algebra basis. We define
\begin{align}
  t^\pm  & = \frac{1}{\sqrt{2}} (t^1 \pm i t^2) \, , & [t^+, t^-] & = -t^3 \, ,  &
[t^\pm, t^3] & = \pm t^\pm \, ,
\end{align}
which implies $i W_\mu =  W^+_\mu t^+ + W_\mu^- t^- + W_\mu^3 t^3$. Next, it is needed to study the first non-trivial term of the electroweak Wilson line $T[w_+]$ 
in Eq.~(\ref{TWilsonLine}):
\begin{eqnarray}
  -gw_+ \frac{1}{\partial_+} w_+ &=& - \frac{g}{2} \left[  w_+  \frac{1}{\partial_+} w_+  - \left(\frac{1}{\partial_+} w_+ \right)  w_+  \right] ~= \frac{g}{2} W_+^a  \left(\frac{1}{\partial_+} W^b_+ \right) \cdot [t^a, t^b] \notag \\
&&\hspace{-2cm}=  \frac{g}{2} \left[ W_+^+   \left(\frac{1}{\partial_+} W^-_+ \right) -  W_+^-   \left(\frac{1}{\partial_+} W^+_+ \right) \right] (-t^3) ~+~  \frac{g}{2} \left[ W_+^+   \left(\frac{1}{\partial_+} W^3_+ \right) -  W_+^3   \left(\frac{1}{\partial_+} W^+_+ \right) \right] t^+ \notag \\
&+& \frac{g}{2} \left[ W_+^3   \left(\frac{1}{\partial_+} W^-_+ \right) -  W_+^-   \left(\frac{1}{\partial_+} W^3_+ \right) \right] t^- \, ,
\end{eqnarray}
where  $W^3_+ = \left(g Z_+  + g' C_+ \right) / \sqrt{g^2 + {g'}^2}
= \cos{(\theta_W)} Z_+ + \sin{(\theta_W)} C_+$. The relevant induced vertices stem from 
\begin{align}
  \mathcal{L}_{ind}^{\mathcal{O}(g)} & = \frac{1}{2} \tr \left\{ -gw_+ \frac{1}{\partial_+} w_+   \left[\partial^2 A_-^{(w)}  +
     M^2 \left( {g^2}  A_-^{(w)} +  {gg'} A_-^{(b)} \right) \right]  \right\}
\notag \\
& +
\frac{1}{2} \tr \left\{  \left[  A_-^{(w)} \partial^2  + \left({g^2} A_-^{(w)}  + {gg'} 
 A_-^{(b)}
\right) M^2\right] \left(  -gw_+ \frac{1}{\partial_+} w_+  \right)  \right\}.
\end{align}
To evaluate the traces it is enough to study the projection of the term in the squared bracket on the three different generators. One finds
\begin{align}
 i  \, \tr \left[ \left\{  \left[  A_-^{(w)} \partial^2  + \left({g^2} A_-^{(w)}  + {gg'} 
 A_-^{(b)}
\right) M^2\right],  t^+ \right\} \right] & =  \left(\partial^2 + M_W^2 \right)   A_-^{{(W^-)}} \, ,\notag \\
 i  \, \tr \left[ \left\{  \left[  A_-^{(w)} \partial^2  + \left({g^2} A_-^{(w)}  + {gg'} 
 A_-^{(b)}
\right) M^2\right],  t^- \right\} \right] & = \left(\partial^2 + M_W^2 \right)   A_-^{{(W^+)}} \, ,\notag \\
i   \, \tr \left[ \left\{  \left[  A_-^{(w)} \partial^2  + \left({g^2} A_-^{(w)}  + {gg'} 
 A_-^{(b)}
\right) M^2\right],  t^3 \right\} \right] & =  \cos{(\theta_W)} \left(\partial^2 + M_Z^2 \right)   A_-^{{(Z)}} + \sin{( \theta_W)}  \partial^2  A_-^{{(\gamma)}} \, .
\end{align}
The remaining task is to write down the corresponding momentum space expressions.

The two induced vertices of interest in our calculations can be depicted as follows, 
\begin{align}
  \label{2vertices}
  \parbox{3.5cm}{\center \includegraphics[height=2.5cm]{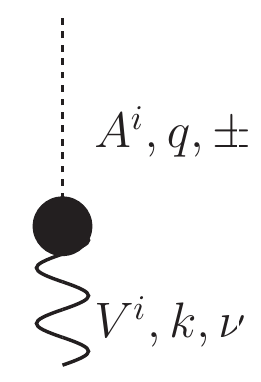}}  & = \frac{i}{2} (q^2 - M_i^2)  (n^\mp)^{\nu} \, ,\notag \\
  \parbox{4cm}{\center \includegraphics[height=2.5cm]{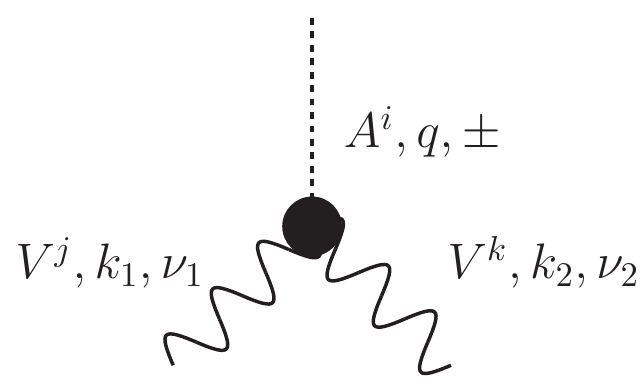}}   & = g K(A^i, V^j, V^k) \cdot \frac{-i}{2 k_1^\mp} (n^\mp)^{\nu_1}(n^\mp)^{\nu_2} \, , \notag 
\end{align}
where $i,j,k  = W^\pm, Z, \gamma,$ and, from the induced Lagrangian, we obtain
 \begin{align}
   K&\left[A_\pm^{W^-}(q), W_{\nu_1}^+(k_1), Z_{\nu_2}(k_2) \right]  =  K\left[A_\pm^{W^+}(q), Z_{\nu_1}(k_1), W^-_{\nu_2}(k_2) \right] = \cos \theta_W ({\bm q}^2 + M_W^2)   \, ,
\notag \\ 
 K&\left[A_\pm^{W^-}(q), W_{\nu_1}^+(k_1), \gamma_{\nu_2}(k_2) \right]  =  K\left[A_\pm^{W^+}(q), \gamma_{\nu_1}(k_1), W^-_{\nu_2}(k_2) \right] =  \sin\theta_W  ({\bm q}^2 + M_W^2)    \, ,
\notag \\ 
   K&\left[A_\pm^{Z}(q), W_{\nu_1}^+(k_1), W^-_{\nu_2}(k_2) \right]  = ({\bm q}^2 + M_Z^2) \cos \theta_W    \, , \notag \\
 K&\left[A_\pm^{\gamma}(q), W_{\nu_1}^+(k_1), W^-_{\nu_2}(k_2) \right]  = {\bm q}^2 \sin \theta_W  \, .
 \end{align}
For clarity, field polarizations and momenta have been written explicitly. 

\section{Reggeization of the electroweak bosons}

In the {\bf charged sector}, the $W^\pm$ bosons are known to lie on their Regge trajectories in the broken symmetry case, {\it i.e.} the Born amplitude for the scattering of $W$ bosons of a given helicity and mass $M_W$ has the all-orders form in the LLA (with $1+ \omega_{\rm charged}(t)$ being the W boson Regge trajectory)
\begin{eqnarray}
A \simeq \frac{A_{\rm Born}}{2} \left(\left(\frac{s}{M_W^2}\right)^{\omega_{\rm charged}(t)}+\left(-\frac{s}{M_W^2}\right)^{\omega_{\rm charged}(t)}\right) \, .
\label{Wamplitude}
\end{eqnarray}  
To calculate the Regge trajectory it is needed to evaluate one-loop corrections to the amplitude for a charged isospin-1 exchange and look for rapidity divergent contributions (which are due to the kinematics, not to short distance dynamics). 
The starting point is the bare propagator in unitary gauge
\begin{align}
 {G}_{\rm charged}^B\left(\rho \right) & = \frac{2i}{{\bm q}^2 + M_W^2}  \left\{1 + \frac{2i}{{\bm q}^2 + M_W^2} \Sigma_{\rm charged}\left(\rho, \epsilon,  {\bm q}^2 \right)  + \ldots \right\} \, ,
\end{align}
where we focus on the high energy divergent piece, {\it i.e.}
\begin{eqnarray}
  \Sigma_{\rm charged}\left(\rho, \epsilon,  {\bm q}^2 \right) &=& -i  K_1^2
 \frac{g^2}{(4 \pi)^{d/2}} \int \frac{d^d l}{i \pi^{d/2}} \frac{n^\mu_+ n^\nu_+n^{\mu'}_+n^{\nu'}_+}{2l^+ 2l^-} G^{\mu\nu}_{W^\pm}(l) G^{\mu'\nu'}_{Z}(k+l)  \notag \\
&-& i  K_2^2 \frac{g^2}{(4 \pi)^{d/2}} \int \frac{d^d l}{i \pi^{d/2}} \frac{n^\mu_+ n^\nu_+n^{\mu'}_+n^{\nu'}_+}{2l^+ 2l^-} G^{\mu\nu}_{W^\pm}(l) G^{\mu'\nu'}_{\gamma}(k+l)  \notag \\
&\simeq &  - \rho \frac{  ({\bm q}^2 + M_W^2)^2}{2i} \left[ \beta_{WZ}({\bm q}^2) \cos^2{(\theta_W)} + \beta_{W\gamma}({\bm q}^2) \sin^2{(\theta_W)} \right] \, ,
\end{eqnarray}
where
\begin{align}
  G^{\mu\nu}_A(l) & = \frac{-i}{l^2 - m_A^2} \left(g^{\mu\nu} - \frac{l^\mu l^\nu}{m_A^2} \right) \, , 
\end{align}
$p = n_+ + e^{-\rho} n_-$, $n = n_- + e^{-\rho} n_+$, $p \cdot n \simeq  2$, $p^2 = n^2 = 4 e^{-\rho}$ and  
\begin{align}
   \rho \simeq \frac{1}{2} \log \left(\frac{ 4 n \cdot p^2}{n^2 p^2}\right) \, .
\end{align}
We have used  the notation of~\cite{Bartels:2006kr} (together with $g^2 = 4 \pi \alpha_{ew}, \bar{\alpha}_{ew} = 2 \alpha_{ew} / \pi$) for 
\begin{align}
  \beta_{ij}({\bm q}^2) & = 2\frac{g^2}{(4\pi)^2} \int \frac{d^2{\bm l}}{\pi} \frac{1}{[{\bm l}^2 + M_i^2] [{\bm l} + {\bm q}^2 + M_j^2]} \, .
\end{align}
We can then write
\begin{align}
 {G}^B_{\rm charged} \left(\rho \right) 
  =  \frac{2i}{{\bm q}^2 + M_W^2} -  2 i \rho   \left[ \beta_{WZ}({\bm q}^2) \cos^2{(\theta_W)} + \beta_{W\gamma}({\bm q}^2) \sin^2{(\theta_W)} \right]
 \, .
\end{align}
In the LLA this expression is enough to show agreement with the results of~\cite{Bartels:2006kr} for the reggeization of the charged electroweak bosons with trajectory
\begin{eqnarray}
\omega_{\rm charged} (t) &=& (t-M_W^2) \left(\beta_{WZ}(-t) \cos^2{(\theta_W)} + \beta_{W\gamma}(-t) \sin^2{(\theta_W)} \right) \, .
\end{eqnarray}
At higher orders it is needed to introduce the renormalization procedure proven to be valid at two loops in the case of the high energy  QCD effective action and put forward in Ref.~\cite{Hentschinski:2008rw,Hentschinski:2011tz,Hentschinski:2011xg,Chachamis:2012gh,Chachamis:2012mw,Chachamis:2012cc,Chachamis:2013hma,Hentschinski:2014lma,Hentschinski:2014bra,Hentschinski:2014esa}.

\vspace{1cm}

The treatment of the {\bf neutral sector}
 is more complicated since the $Z$ and $\gamma$ bosons do not reggeize within the Stantard Model (they do in grand unified theories~\cite{Grisaru:1979wi,Lukaszuk:1979zj}) and 
 a representation such as Eq.~(\ref{Wamplitude}) does not apply. In this case both the $Z$ and 
$\gamma$ high energy fields can split  into a $W^+ W^-$ pair and mix in the $t$-channel. This fact generates four possible bare propagators which at one loop, and keeping only the high energy divergent contributions (proportional to $\rho \sim \ln{s}$), can be written (using the same notations as above) in the form
\begin{align}
 \widehat{G}_{\rm neutral}^B (\rho) = \frac{1}{2i}
              \begin{pmatrix}
                {G}_{ZZ}^B & {G}_{Z\gamma}^B \\
                   {G}_{\gamma Z}^B & {G}_{\gamma\gamma}^B
              \end{pmatrix} =  \widehat{G}_0 - \rho \, \beta_{WW}({\bm q}^2) 
              \,  \widehat{G}_1 \, ,
\end{align} 
where
\begin{align}
 & \widehat{G}_0 =   \widehat{G}_{\rm neutral}^B (0) = 
                                         \begin{pmatrix}
                                           \frac{1}{{\bm q}^2 + M_Z^2} &   0 \\
                                           0 &  \frac{1}{{\bm q}^2}
                                         \end{pmatrix} \, ,           
\end{align} 
and 
\begin{eqnarray}
 \widehat{G}_1 =       \begin{pmatrix}
                        \cos^2{(\theta_W)} & \frac{\sin{(2 \theta_W)}}{2} \\
                         \frac{\sin{(2 \theta_W)}}{2} & \sin^2{(\theta_W)}
                  \end{pmatrix}    = 
                  \frac{{\bm q}^2 + M_W^2}{2} \left[
 \widehat{S} \begin{pmatrix}
                                           1&   0 \\
                                           0 &  0
                                         \end{pmatrix}
                                          \widehat{S}^{-1} \widehat{G}_0 
                                       +  \widehat{G}_0 \left(\widehat{S}^{-1}\right)^T \begin{pmatrix}
                                           1&   0 \\
                                           0 &  0
                                         \end{pmatrix}
                                          \widehat{S}^{T}  \right] \, ,
\end{eqnarray}
with
\begin{align}
& \widehat{S}  =   \begin{pmatrix}
                                           \cos{(\theta_W)}&   - \frac{{\bm q}^2}{{\bm q}^2 + M_Z^2}\sin{(\theta_W)}  \\
                                           \sin{(\theta_W)} &  \cos{(\theta_W)}
                                         \end{pmatrix}    \, ,                                                
                                         & \widehat{S}^{-1}  =   
                                         \frac{{\bm q}^2 + M_Z^2 }{{\bm q}^2 + M_W^2 }
                                         \begin{pmatrix}
                                           \cos{(\theta_W)}&   \frac{{\bm q}^2}{{\bm q}^2 + M_Z^2}\sin{(\theta_W)}  \\
                                           -\sin{(\theta_W)} &  \cos{(\theta_W)}
                                         \end{pmatrix} \, .                                                             
\end{align}
This result is in agreement with those presented in~\cite{Bartels:2006kr} where there exists a Regge trajectory, 
\begin{eqnarray}
\omega_{\rm neutral} (t) = (t-M_W^2)\beta_{WW} (-t) \, ,
\end{eqnarray}
in the neutral sector. However, this trajectory $\alpha_{\rm neutral} (t) = 1+ \omega_{\rm neutral} (t)$ crosses one at $t=M_W^2$ which implies that the neither the Z boson nor the photon lie on it.

\section{Emission vertices}

The second ingredient needed to construct the integral equations governing the 
high energy behaviour of electroweak scattering amplitudes in the multi-Regge limit 
is the set of real emission vertices with electroweak bosons or a Higgs in the final state. 

\begin{figure}[th]
  \centering
  \parbox{2cm}{\includegraphics[height=3cm]{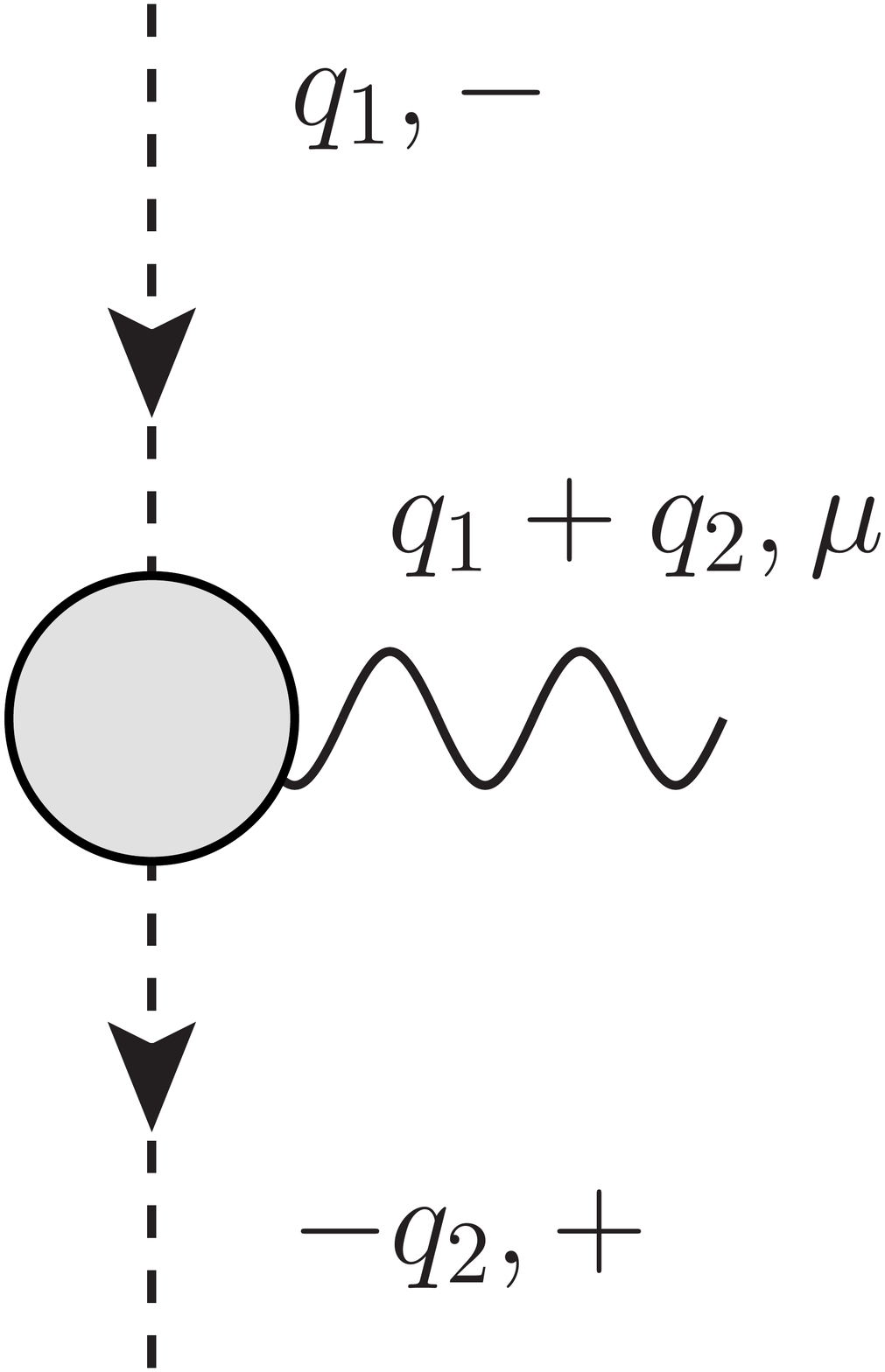}} 
  =
    \parbox{1.7cm}{\includegraphics[height=3cm]{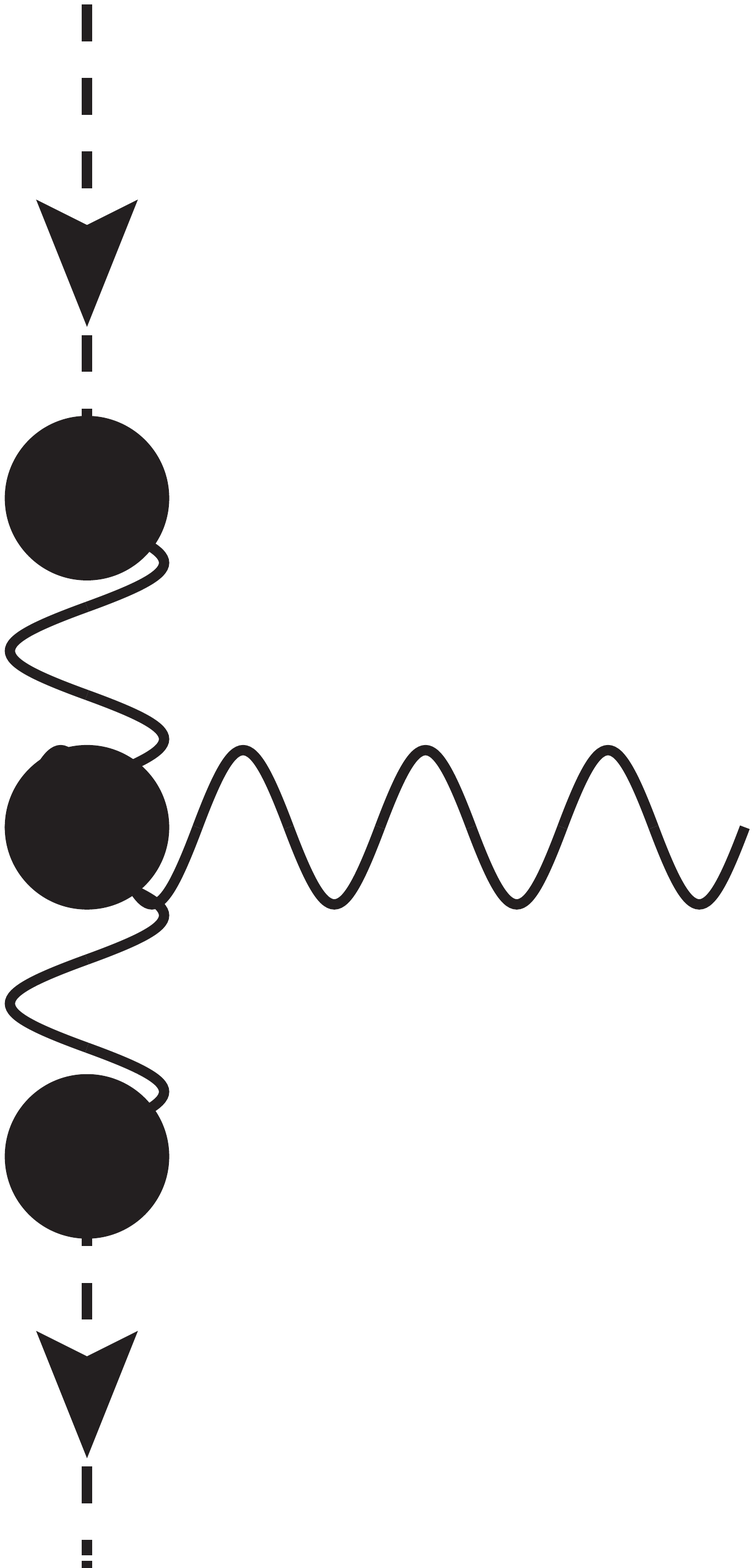}}
    +
    \parbox{1.7cm}{\includegraphics[height=3cm]{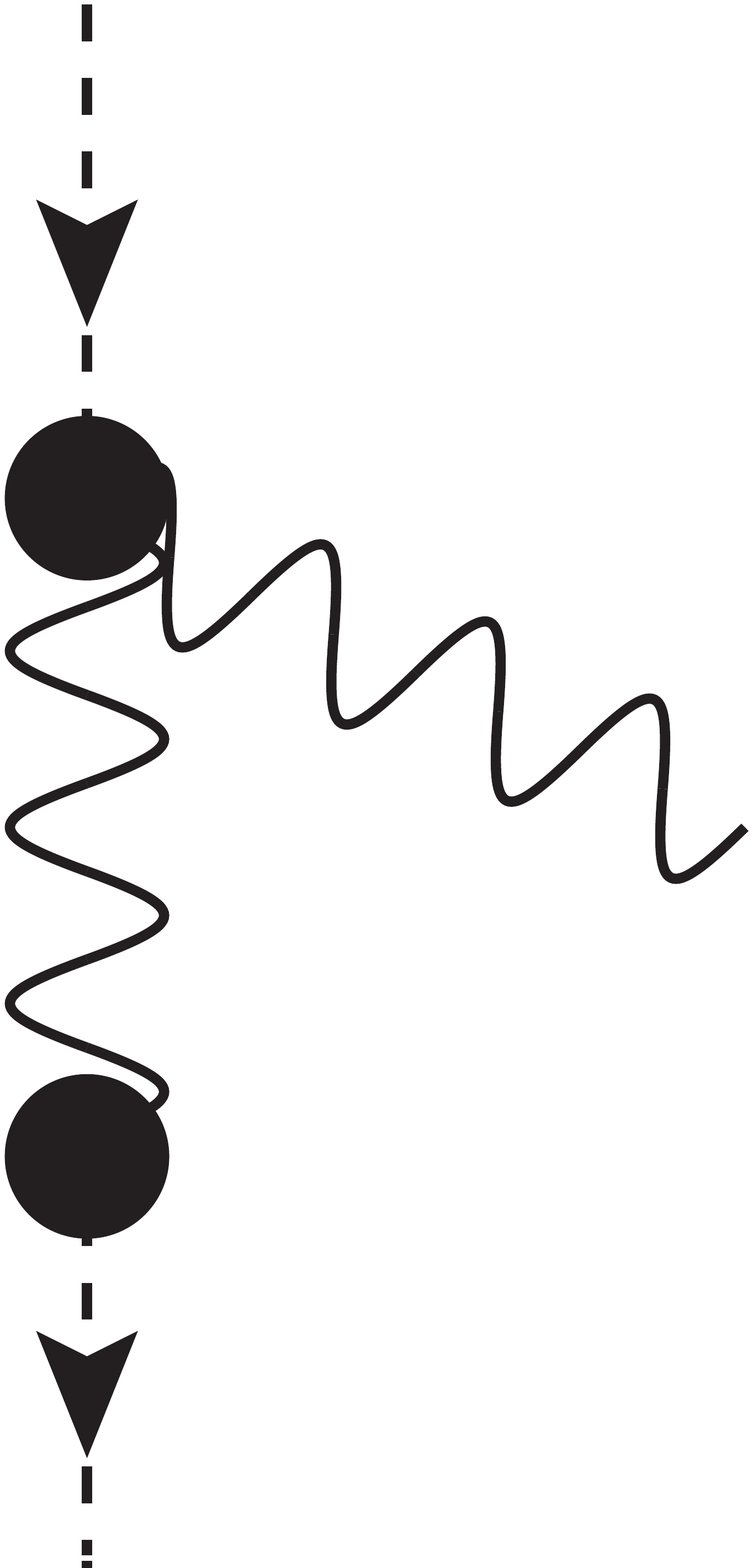}}
    +
    \parbox{1.7cm}{\includegraphics[height=3cm]{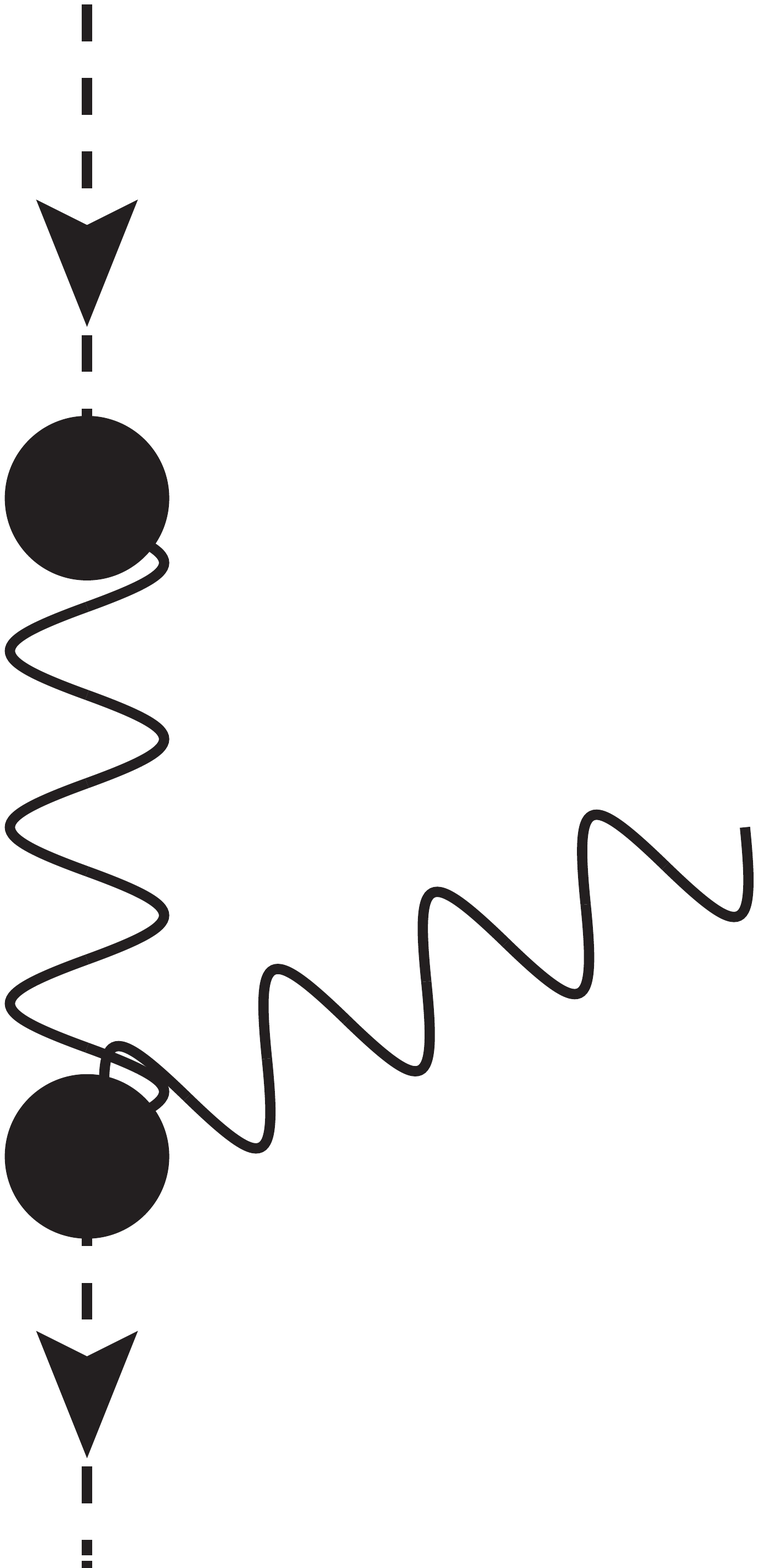}} \hspace{2cm}
 \parbox{2cm}{\includegraphics[height=3cm]{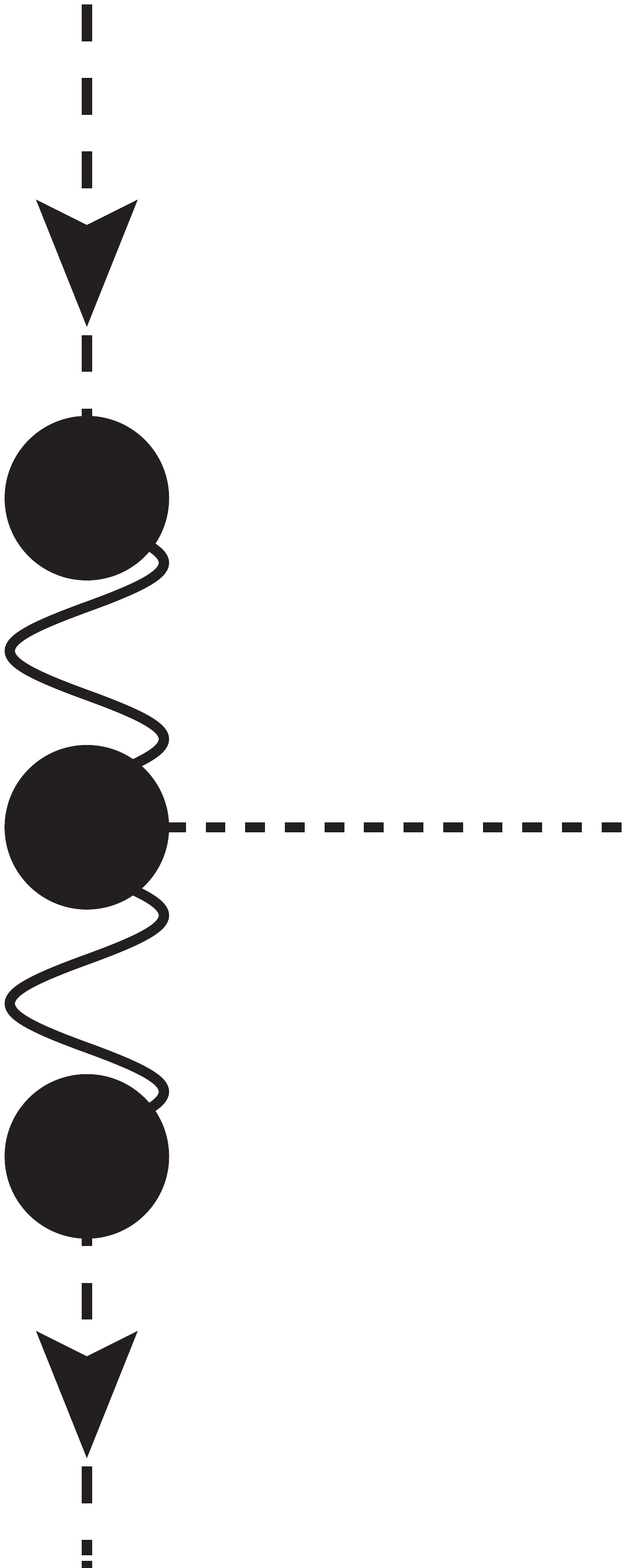}}
  \caption{Production of a photon or $Z$ boson (left) and a Higgs boson (right) from a charged $W$ vector boson}
  \label{prod_fromW}
\end{figure}

The effective action proposed in this work allows to obtain the same central rapidity 
production vertices as those calculated by Bartels, Lipatov and Peters in the LLA in Ref.~\cite{Bartels:2006kr}. 
In the case of the exchange of a high energy $W$ boson in the $t$-channel with associated production of a photon the three Feynman diagrams in the effective field theory are shown in Fig.~\ref{prod_fromW} (left). 

The sum of these diagrams renders the expression for the vertex
\begin{align}
  C_\mu^{W^+W^-\gamma} & = \frac{ie}{2} \left[-q_1^+ n_\mu^- + q_2^- n_\mu^+ +  (q_1 - q_2)^\mu \right] - \frac{ie}{2} \frac{q_1^2 - M_W^2}{q_2^-} n_\mu^- 
+ \frac{ie}{2} \frac{q_2^2 - M_W^2}{q_1^+} n_\mu^+ \, .
\end{align}
There is a one-to-one correspondence between the three terms in this vertex and each of the graphs in the figure. 

The corresponding vertex for the exchange of a $Z$ boson can be calculated by making the replacement  $e \to g \cos{(\theta_W)} =  e \cot{(\theta_W)}$. 

The graph for Higgs production shown in Fig.~\ref{prod_fromW} (right) corresponds to 
the vertex
\begin{align}
  C^{W^+W^-H} & = 2i \frac{m_W^2}{v} \, .
\end{align}

The remaining production vertices are calculated with the same method.  Let us fix our notation for the above mentioned vertex $C_\mu^{W^+W^-\gamma}$ (see Fig.~\ref{prod_fromW} (left)) as $r_+^{W^+}(q_1) + r_-^{W^-} (q_2)\to \gamma (q_1 + q_2)$, etc. (where ``$r_\pm^A (q)$" stands for a Reggeon  $A$ exchanged in the $t$-channel with + (-) lightcone projection and transverse momentum $q$). We then find for $r_+^{\gamma}(q_1) + r_-^{W^+} (q_2)\to W^+ (q_1 + q_2) $:
\begin{align}
  C_\mu^{\gamma W^+W^+} & = \frac{ie}{2} \left[q_1^+ n_\mu^- - q_2^- n_\mu^+ +  (q_2 - q_1)^\mu \right] + \frac{ie}{2} \frac{q_1^2}{q_2^-}n_\mu^- -\frac{ie}{2} \frac{q_2^2 - M_W^2}{q_1^+} n_\mu^+ \, ,
\end{align}
and for $r_+^{Z}(q_1) + r_-^{W^+} (q_2)\to W^+ (q_1 + q_2) $:
\begin{align}
  \hspace{-.2cm} C_\mu^{Z W^+W^+} \hspace{-.3cm}=  \hspace{-.1cm}\frac{ig \cos \theta_W}{2} \left[q_1^+ n_\mu^- - q_2^- n_\mu^+ +  (q_2 - q_1)^\mu \right] 
   \hspace{-.1cm}+  \hspace{-.1cm}\frac{ig \cos \theta_W}{2} \frac{q_1^2 - M_Z^2}{q_2^-} n_\mu^- 
    \hspace{-.1cm}- \hspace{-.1cm}\frac{ig \cos \theta_W}{2} \frac{q_2^2 - M_W^2}{q_1^+} n_\mu^+ . \hspace{-.25cm}
\end{align}
These vertices are equivalent to those calculated in~\cite{Bartels:2006kr}. This concludes our analysis of the effective action  in the LLA since with the results of the last two sections it is possible to reproduce, using the product of pairs of these effective vertices and applying $s$-channel unitarity, both the vacuum and isospin-1 exchange integral equations derived in~\cite{Bartels:2006kr}. The detailed study of the solutions to those equations (see~\cite{Bartels:2016ejz} for some studies in this direction) and the application of this formalism at NLL will be defered to a future work.

\section{Conclusions}

We have introduced an effective action valid to evaluate scattering amplitudes in the high energy limit of the electroweak sector of the Standard Model. We have made use of Wilson lines to deal with reggeized charged electroweak vector bosons and investigated their effective interactions among themselves and with the non-reggeizing neutral vector bosons. We have found agreement with the results of~\cite{Bartels:2006kr} for the charged electroweak vector boson Regge trajectories and production vertices needed to write the evolution equations governing the singlet and isospin-1 exchanges in the $t$-channel in multi-Regge kinematics. The results here presented belong to the leading logarithmic approximation but the proposed high energy effective action is valid beyond this. We will evaluate the next-to-leading order corrections to these results in future works making use of renormalized high energy propagators. This method has already been proven to be efficient at that order in the QCD case~\cite{Chachamis:2013hma}. Probably the most interesting feature of the next-to-leading order calculation is to test if the reggeization in the neutral sector could already  appear at that level of accuracy. In the effective theory context this 
would take place in the renormalization procedure mandatory for the associated propagators. A further interesting aspect of the 
calculation is to evaluate the likely presence of double logarithms 
related to the choice of energy scale in the resummed terms. The 
identification of these terms will, most likely, lead to their resummation 
to all orders using very similar arguments to those in~\cite{Vera:2005jt}.

\section*{Acknowledgements}

Support by Consejo Nacional de Ciencia y Tecnolog{\'\i}a grant number A1 S-43940 (CONACYT-SEP Ciencias B{\'a}sicas) is gratefully acknowledged. This work has been supported by the Spanish Research Agency (Agencia Estatal de Investigaci\'on) through the grant IFT Centro de Excelencia Severo Ochoa SEV-2016-0597, and the Spanish Government grant FPA2016-78022-P.  It has also received
funding from the European Union’s Horizon 2020 research
and innovation programme under grant agreement No.
824093.


\begin{thebibliography}{99}

\bibitem{Bartels:2006kr}
J.~Bartels, L.~N.~Lipatov, K.~Peters,
Nucl. Phys. B \textbf{772} (2007), 103-132.

  
\bibitem{Lipatov:1976zz}
  L.~N.~Lipatov,
  Sov.\ J.\ Nucl.\ Phys.\  {\bf 23} (1976) 338
   [Yad.\ Fiz.\  {\bf 23} (1976) 642].
  
\bibitem{Fadin:1975cb}
  V.~S.~Fadin, E.~A.~Kuraev, L.~N.~Lipatov,
  Phys.\ Lett.\  {\bf 60B} (1975) 50.
  
\bibitem{Kuraev:1976ge}
  E.~A.~Kuraev, L.~N.~Lipatov, V.~S.~Fadin,
  Sov.\ Phys.\ JETP {\bf 44} (1976) 443
   [Zh.\ Eksp.\ Teor.\ Fiz.\  {\bf 71} (1976) 840].
  
\bibitem{Kuraev:1977fs}
  E.~A.~Kuraev, L.~N.~Lipatov, V.~S.~Fadin,
  Sov.\ Phys.\ JETP {\bf 45} (1977) 199
   [Zh.\ Eksp.\ Teor.\ Fiz.\  {\bf 72} (1977) 377].
  
\bibitem{Balitsky:1978ic}
  I.~I.~Balitsky, L.~N.~Lipatov,
  Sov.\ J.\ Nucl.\ Phys.\  {\bf 28} (1978) 822
   [Yad.\ Fiz.\  {\bf 28} (1978) 1597].
  
\bibitem{Balitsky:1979ns}
  I.~I.~Balitsky, L.~N.~Lipatov,
  JETP Lett.\  {\bf 30} (1979) 355
   [Pisma Zh.\ Eksp.\ Teor.\ Fiz.\  {\bf 30} (1979) 383].
  
\bibitem{Lipatov:1995pn}
  L.~N.~Lipatov,
  Nucl.\ Phys.\ B {\bf 452} (1995) 369.
  
\bibitem{Lipatov:1996ts}
  L.~N.~Lipatov,
  Phys.\ Rept.\  {\bf 286} (1997) 131.
  
\bibitem{Antonov:2004hh}
  E.~N.~Antonov, L.~N.~Lipatov, E.~A.~Kuraev, I.~O.~Cherednikov,
  Nucl.\ Phys.\ B {\bf 721} (2005) 111.
  
  



\bibitem{Hentschinski:2008rw}
  M.~Hentschinski,
  Acta Phys.\ Polon.\ B {\bf 39} (2008) 2567.
  
\bibitem{Hentschinski:2011tz}
  M.~Hentschinski, A.~Sabio Vera,
  Phys.\ Rev.\ D {\bf 85} (2012) 056006.
  
\bibitem{Hentschinski:2011xg}
  M.~Hentschinski,
  Nucl.\ Phys.\ B {\bf 859} (2012) 129.
  
\bibitem{Chachamis:2012gh}
  G.~Chachamis, M.~Hentschinski, J.~D.~Madrigal Martinez, A.~Sabio Vera,
  Nucl.\ Phys.\ B {\bf 861} (2012) 133.
  
\bibitem{Chachamis:2012mw}
  G.~Chachamis, M.~Hentschinski, J.~D.~Madrigal Martínez, A.~Sabio Vera,
  Phys.\ Part.\ Nucl.\  {\bf 45} (2014) no.4,  788.
  
\bibitem{Chachamis:2012cc}
  G.~Chachamis, M.~Hentschinski, J.~D.~Madrigal Martínez, A.~Sabio Vera,
  Phys.\ Rev.\ D {\bf 87} (2013) no.7,  076009.
  
\bibitem{Chachamis:2013hma}
  G.~Chachamis, M.~Hentschinski, J.~D.~Madrigal Martinez, A.~Sabio Vera,
  Nucl.\ Phys.\ B {\bf 876} (2013) 453.
  
\bibitem{Hentschinski:2014lma}
  M.~Hentschinski, J.~D.~Madrigal Martínez, B.~Murdaca, A.~Sabio Vera,
  Phys.\ Lett.\ B {\bf 735} (2014) 168.
  
\bibitem{Hentschinski:2014bra}
  M.~Hentschinski, J.~D.~Madrigal Martínez, B.~Murdaca, A.~Sabio Vera,
  Nucl.\ Phys.\ B {\bf 887} (2014) 309.
  
\bibitem{Hentschinski:2014esa}
  M.~Hentschinski, J.~D.~M.~Martínez, B.~Murdaca, A.~Sabio Vera,
  Nucl.\ Phys.\ B {\bf 889} (2014) 549.
   

\bibitem{Ciafaloni:2008cr}
  M.~Ciafaloni, P.~Ciafaloni, D.~Comelli,
  JHEP {\bf 0805} (2008) 039.
  
\bibitem{Bauer:2017isx}
  C.~W.~Bauer, N.~Ferland, B.~R.~Webber,
  JHEP {\bf 1708} (2017) 036.
  
\bibitem{Bauer:2017bnh}
  C.~W.~Bauer, N.~Ferland, B.~R.~Webber,
  JHEP {\bf 1804} (2018) 125.
  
\bibitem{Bauer:2018xag}
  C.~W.~Bauer, D.~Provasoli, B.~R.~Webber,
  JHEP {\bf 1811} (2018) 030.
  
\bibitem{Manohar:2018kfx}
  A.~V.~Manohar, W.~J.~Waalewijn,
  JHEP {\bf 1808} (2018) 137.
  
\bibitem{Fornal:2018znf}
  B.~Fornal, A.~V.~Manohar, W.~J.~Waalewijn,
  JHEP {\bf 1805} (2018) 106.
  
\bibitem{Nefedov:2013ywa}
  M.~A.~Nefedov, V.~A.~Saleev, A.~V.~Shipilova,
  Phys.\ Rev.\ D {\bf 87} (2013) no.9,  094030.
  
\bibitem{Bondarenko:2018kqs}
  S.~Bondarenko, S.~Pozdnyakov,
  Phys.\ Lett.\ B {\bf 783} (2018) 207.
  
\bibitem{Hentschinski:2018rrf}
  M.~Hentschinski,
  Phys.\ Rev.\ D {\bf 97} (2018) no.11,  114027.
  
\bibitem{Karpishkov:2017kph}
  A.~V.~Karpishkov, M.~A.~Nefedov, V.~A.~Saleev,
  Phys.\ Rev.\ D {\bf 96} (2017) no.9,  096019.
  
\bibitem{Braun:2017qij}
  M.~A.~Braun, M.~Y.~Salykin,
  Eur.\ Phys.\ J.\ C {\bf 77} (2017) no.7,  498.
  
\bibitem{Braun:2016sij}
  M.~A.~Braun, M.~I.~Vyazovsky,
  Phys.\ Rev.\ D {\bf 93} (2016) no.6,  065026.
  
\bibitem{Nefedov:2014qea}
  A.~V.~Karpishkov, M.~A.~Nefedov, V.~A.~Saleev, A.~V.~Shipilova,
  Phys.\ Rev.\ D {\bf 91} (2015) no.5,  054009.
  
\bibitem{Nefedov:2020ecb}
  M.~A.~Nefedov,
  JHEP {\bf 2008} (2020) 055.
  
\bibitem{Bondarenko:2019llt}
  S.~Bondarenko, S.~Pozdnyakov,
  Nucl.\ Phys.\ B {\bf 951} (2020) 114854.
  
\bibitem{Braun:2020thc}
  M.~A.~Braun,
  Eur.\ Phys.\ J.\ C {\bf 80} (2020) no.8,  774.
  
\bibitem{Bartels:2008ce}
  J.~Bartels, L.~N.~Lipatov and A.~Sabio Vera,
  Phys.\ Rev.\ D {\bf 80} (2009) 045002.
  
\bibitem{Bartels:2008sc}
  J.~Bartels, L.~N.~Lipatov, A.~Sabio Vera,
  Eur.\ Phys.\ J.\ C {\bf 65} (2010) 587.
  
\bibitem{Lipatov:1982vv}
  L.~N.~Lipatov,
  Phys.\ Lett.\  {\bf 116B} (1982) 411.

\bibitem{Lipatov:1991nf}
  L.~N.~Lipatov,
  Nucl.\ Phys.\ B {\bf 365} (1991) 614.
  
\bibitem{Lipatov:2011ab}
  L.~N.~Lipatov,
  Phys.\ Part.\ Nucl.\  {\bf 44} (2013) 391.
  
\bibitem{Bartels:2012ra}
  J.~Bartels, L.~N.~Lipatov, A.~Sabio Vera,
  JHEP {\bf 1407} (2014) 056.
  
\bibitem{SabioVera:2019edr}
  A.~Sabio Vera,
  JHEP {\bf 1907} (2019) 080.
  
\bibitem{SabioVera:2019jqe}
  A.~Sabio Vera,
  JHEP {\bf 2001} (2020) 163.
  
\bibitem{Grisaru:1973vw}
  M.~T.~Grisaru, H.~J.~Schnitzer, H.~S.~Tsao,
  Phys.\ Rev.\ Lett.\  {\bf 30} (1973) 811.

  
\bibitem{Grisaru:1979wi}
  M.~T.~Grisaru, H.~J.~Schnitzer,
  Phys.\ Rev.\ D {\bf 20} (1979) 784.
  
\bibitem{Lukaszuk:1979zj}
  L.~Lukaszuk, L.~Szymanowski,
  Nucl.\ Phys.\ B {\bf 159} (1979) 316.

\bibitem{Bartels:2016ejz}
  J.~Bartels, E.~Levin, M.~Siddikov,
  Phys.\ Rev.\ D {\bf 94} (2016) no.5,  053012.
  
\bibitem{Vera:2005jt}
  A.~Sabio Vera,
  Nucl.\ Phys.\ B {\bf 722} (2005) 65.


\end{thebibliography}
\end{document}